\documentclass[aps,prl,twocolumn,superscriptaddress]{revtex4}

\usepackage{amsfonts,epsfig}
\usepackage{latexsym}
\usepackage{epsfig}
\usepackage{amsmath}
\usepackage{amssymb}
\usepackage{amsfonts}
\usepackage{amsthm}
\usepackage{mathrsfs}
\usepackage{natbib}
\usepackage{color,verbatim,graphics}
\usepackage{psfrag}
\DeclareMathAlphabet{\mathrsfs}{U}{rsfs}{m}{n}
\DeclareMathAlphabet{\mathpzc}{OT1}{pzc}{m}{it}
\DeclareMathAlphabet{\matheus}{U}{eus}{m}{n}
\DeclareMathAlphabet{\mathbbold}{U}{bbold}{m}{n}

\newcommand{\ba}{\begin{eqnarray}}
\newcommand{\be}{\begin{equation}}
\newcommand{\ee}{\end{equation}}

\newcommand{\ea}{\end{eqnarray}}
\newcommand{\ban}{\begin{eqnarray*}}
\newcommand{\ean}{\end{eqnarray*}}
\newcommand{\ket}[1]{|#1\rangle}
\newcommand{\bra}[1]{\langle#1|}

\newcommand{\eg}{{\it{e.g.}}}
\newcommand{\ie}{{\it{i.e.}}}
\newcommand{\etal}{{\it{et al.}}}

\begin{document}

%\title{Activation and noise robustness of nonlocality in quantum networks}
\title{Nonlocality tests enhanced by a third observer}

\author{Daniel Cavalcanti}
%\email{dcavalcanti@nus.edu.sg}
\affiliation{Centre for Quantum Technologies, National University of Singapore, 3 Science Drive 2, Singapore 117543}%
\author{Rafael Rabelo}
\affiliation{Centre for Quantum Technologies, National University of Singapore, 3 Science Drive 2, Singapore 117543}%
\author{Valerio Scarani}
\affiliation{Centre for Quantum Technologies, National University of Singapore, 3 Science Drive 2, Singapore 117543}
\affiliation{Department of Physics, National University of Singapore, 2 Science Drive 3, Singapore 117542}

%etc.

%\date{\today}

%______________________________________________________________________ ABSTRACT

\begin{abstract}
We consider Bell tests involving bipartite states shared between three parties. 
We show that the simple inclusion of a third part may greatly simplify the measurement scenario (in terms of the number of measurement settings per part) and allows the identification
of previously unknown nonlocal resources.
\end{abstract}

\maketitle

\textit{Introduction.--} The implementation of quantum networks is a crucial goal of the quantum communication program \cite{network}. Ultimately, a quantum network aims at distributing quantum correlations among distant locations through necessarily imperfect quantum channels. These correlations can later be used to perform quantum information and communication protocols \cite{nielsen}. Nonlocal correlations, in the sense of Bell  \cite{bell}, are the prototypical example of quantum correlations, but not all quantum correlations are non-local: famously, there exist states that do not violate any Bell inequality, but which are entangled and in fact distillable \cite{werneretal}. Recently, it has been noticed that nonlocal correlations allow device-independent quantum information protocols (\eg, quantum key distribution \cite{diqkd} and random number generation \cite{dirng}), in which the success can be assessed without the need of characterizing the states and measurements being used. If one wants to implement a device-independent protocol, it is not enough to distribute entanglement: the network must generate nonlocal correlations. This applied motivation adds to the more speculative one of performing fundamental tests of quantum physics at large distances.

We consider nonlocality tests in tripartite networks. We show
the existence of bipartite states $\rho$ from which only local (classical) correlations can be obtained if one of the parties performs a finite number of measurements; however, if two copies of $\rho$ are shared between three parties, nonlocal correlations can be obtained from only two dichotomic observables per party. We also show that high dimensional maximally entangled states can stand an arbitrary amount of separable noise and still be used to distribute tripartite nonlocal correlations. Furthermore, we identify two-qubit states that do not violate any two-input-two-output Bell inequality, but several copies of them do. Our findings show that the simple addition of a third party makes Bell tests much more powerful. 

\textit{Methods.--}
Our results make use of the following observation \cite{pr}. Consider an initial N partite quantum state: if there exist local measurements in $k$ parties such that, for at least one of the measurement outcomes, the remaining $N-k$ parties are projected in a nonlocal state, then the initial state is necessarily nonlocal. This fact can be proved by contradiction: if the initial state is local, any reduced conditional state will also be local.
\begin{figure}
\centering
\includegraphics[width = 0.47\textwidth]{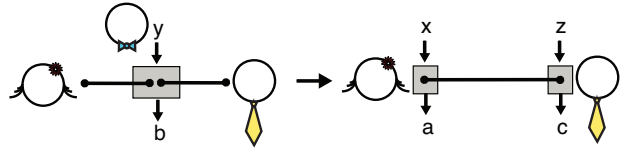}
	\caption{(Color online) Illustration of the methods employed in this letter. Alice, Bob and Charlie firstly share two copies of a bipartite state. Bob, in the middle, performs a measurement $y$ on his subsystem, which, for a given outcome $b$ produces a nonlocal state between Alice and Charlie.}
\label{Fig1A}
\end{figure} 

The results we present are obtained by considering \textit{networks composed of only three parties}. Alice, Bob and Charlie share two copies of a bipartite state $\rho$. To reveal the nonlocality of this tripartite state, Bob applies a measurement aiming to leave Alice and Charlie with a nonlocal state (Fig.~\ref{Fig1A}) \cite{polish}.
Note that Bob's measurement does not need to be a single projective measurement: in fact, the cases of interest below involve the preparation of an ancillary system or a sequence of measurements. In the former case, Bob can prepare a bipartite state $\ket{\Phi}$ and teleport it through the bipartite states composing the network, which can be seen as channels from which the state $\ket{\Phi}$ is distributed. Thus, the most natural choice of $\ket{\Phi}$ is the most robust state against the noise channels defined by the network states. This simple argument allows us to link the problem of deciding if a given state $\rho$ is a nonlocal resource to the problem of finding robust states $\ket{\Phi}$ against the channel defined by $\rho$. This also provides a connection to the problem of finding unbounded violations of Bell inequalities \cite{unbounded BI}.

\textit{Isotropic state.--} First we consider the \textit{isotropic state} \be\label{isotropic}
\rho^{iso}_{AB}=p\ket{\Psi_+^d}\bra{\Psi_+^d}+(1-p)\frac{\openone}{d^2},
\ee
where $\ket{\Psi_+^d}=\sum_{i=0}^{d-1}\ket{ii}/\sqrt{d}$ is the maximally entangled state and $\openone/d^2$ the maximally mixed state, both in $\mathbb{C}^d\otimes\mathbb{C}^d$. This state appears naturally in the scenario where half of a maximally entangled state is sent through a \textit{depolarizing channel} with depolarizing probability $p$.
A single copy of an isotropic state is known to be local for $p\lesssim O(\frac{\log d}{d})$ \cite{noise} (in case of projective measurements) and nonlocal for $p\gtrsim 0.67$, in the limit $d\rightarrow \infty$ \cite{cglmp}. Moreover, the isotropic state was previously shown to be a nonlocal resource for $p>1/2$, again in the limit $d\rightarrow \infty$ \cite{us}. Here we improve this bound and show that the isotropic state is a nonlocal resource for $p> O(\frac{\sqrt{\log d}}{d^{1/4}})$ in the same limit.

We use the fact that there exist bipartite states $\ket{\Phi}$ with local dimension $d$ that achieve unbounded violations of a Bell inequality with respect to $d$ \cite{unbounded BI,unbounded BI 2}. In fact, by measuring such states in the appropriate measurement basis, one obtains a probability distribution $P_\Phi$ such that the distribution
%\be
$qP_\Phi+(1-q)P_{loc}$
%\ee
is nonlocal for $q\geq O(\frac{\log d}{d^{1/2}})$ and any local probability distribution $P_{loc}$, in the limit $d\rightarrow\infty$ \cite{unbounded BI 2}. 

\begin{figure}
\centering
\includegraphics[width = 0.47\textwidth]{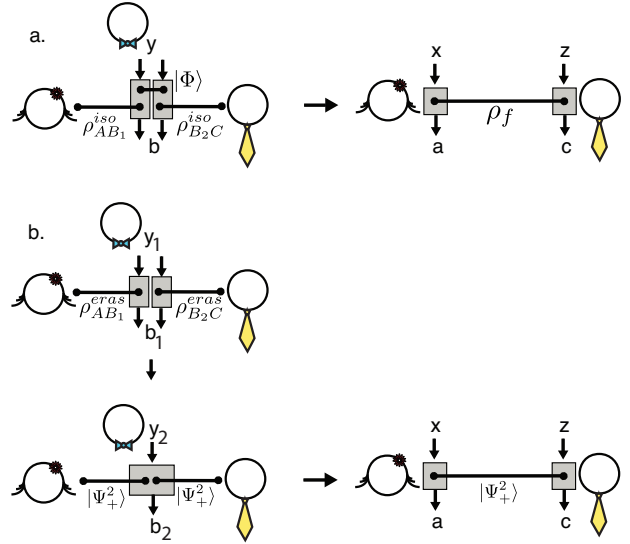}
	\caption{(Color online) Measurement protocol used to obtain nonlocality from the isotropic (Eq. \eqref{isotropic}) and erased (Eq. \eqref{erased}) states. a) Isotropic state: the measurement $y$ consists on preparing a bipartite system in state $\ket{\Phi}$ and then teleporting each of its parts. b) Erased state: The measurement $y$ consists on two steps: first, independent measurements $y_1$ are performed on subsystems $B_1$ and $B_2$; second, a Bell state measurement $y_2$ is performed. }
\label{Fig1BC}
\end{figure}

Consider then the scenario described in Fig.~\ref{Fig1BC}a. Initially, the state $\rho^{iso}_{AB_1} \otimes \rho^{iso}_{B_2 C}$ is shared between Bob, who possesses systems $B_1$ and $B_2$, and Alice and Charlie. Bob performs a joint generalized measurement on his subsystems that corresponds to preparing the state $\ket{\Phi}$ and teleporting each of its components to Alice and Charlie throught the channels defined by the states $\rho^{iso}_{AB_1}$ and $\rho^{iso}_{B_2 C}$. In the case Bob obtains the outcomes corresponding to the state $\ket{\Psi_+^2}$ for each teleportation measurement, the final state shared between Alice and Charlie is
\ba
\rho^{f} &=& p^{2}\ket{\Phi}\bra{\Phi} + p(1-p)\sigma_{A} \otimes \frac{\openone_{C}}{d}\nonumber\\&+& p(1-p)\frac{\openone_{A}}{d} \otimes \sigma_{C} + (1-p)^{2} \frac{\openone_{A}}{d}\otimes\frac{\openone_{C}}{d},
\ea
where $\sigma_{i}$ is the reduced state of part $i$. Performing then appropriate measurements \cite{unbounded BI}, Alice and Charlie end up with a probability distribution of the following form
\be\label{dist}
p^{2}P_{\Phi} + (1-p^2)P_{loc}.
\ee
This distribution, as previously stated, is nonlocal for $p^{2} \geq O(\frac{\log d}{d^{1/2}})$. Thus we conclude that the isotropic state \eqref{isotropic} is a nonlocal resource for $p \geq O(\frac{\sqrt{\log d}}{d^{1/4}})$. Note that this is also valid if we change the state $\openone/d$ in \eqref{isotropic} by any local - in special, separable - state, since it would also result in a distribution like \eqref{dist}.

\begin{figure}
\centering
\includegraphics[width = 0.4\textwidth]{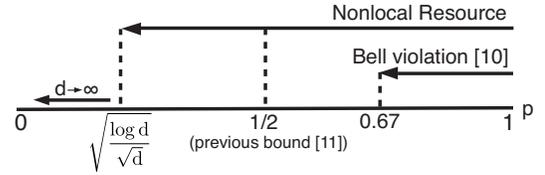}
	\caption{Nonlocality properties of the isotropic state.}
\label{FigIsoBound}
\end{figure}

\textit{Erased state.--} Next we consider the \emph{erased state}, given by
\be\label{erased}
\rho_{AB}^{eras}=\frac{1}{k}\ket{\Psi_+^2}\bra{\Psi_+^2}_{AB}+(1-\frac{1}{k})\frac{\openone_A}{2}\otimes \ket{2}\bra{2}_B,
\ee
where $\ket{\Psi_+^2}=(\ket{00}+\ket{11})/\sqrt{2}$. This is a qubit-qutrit state that can be seen as the result of sending half of a two-qubit maximally entangled state through the \textit{erasure channel} \cite{erasureC}. This channel leaves the state untouched with probability $1/k$ and ``erases'' it with probability $1-1/k$. It is not known whether the erased state is local; however, it has a $k$-symmetric extension with respect to subsystem B, meaning that there exists a state of $k+1$ parties $\rho_{A B_1 B_2 ...B_k}$ such that $\rho_{A B_i}=\rho^{eras}_{AB}$ for every $i$. This implies that only local correlations can be extracted from \eqref{erased} in any experiment where Bob chooses $k$ measurements, regardless of the number of outcomes of those measurements, and regardless of the number of Alice's measurements and their outcomes \cite{terhal}.

We now show that two copies of $\rho_{AB}^{eras}$ shared between three parties violate a Bell inequality where two of the parties perform only two measurements, and one of them performs a single measurement. The state to be considered is $\rho_{AB_1}^{eras}\otimes \rho_{B_2 C}^{eras}$. This state corresponds to two erased states being shared between Alice, Bob, who possesses systems $B_1$ and $B_2$, and Charlie. Moreover Bob carries the qutrit part of the states (\ie, the part whose the extension is possible). 

The measurement protocol that reveals the nonlocality of $\rho_{AB_1}^{eras}\otimes \rho_{B_2 C}^{eras}$ is illustrated in Fig.~\ref{Fig1BC}b. Bob measures both subsystems $B_1$ and $B_2$ with the projective measurement $y_1=\{M_B^0\equiv\ket{0}\bra{0}+\ket{1}\bra{1},M_B^1=\ket{2}\bra{2}\}$. In case he gets outcomes corresponding to the measurement operator $M_B^0$, \ie, both systems $B_1$ and $B_2$ are projected in the subspace $\ket{0}\bra{0}+\ket{1}\bra{1}$, the global state is projected in the state $\ket{\Psi_+^2}\bra{\Psi_+^2}_{AB_1}\otimes\ket{\Psi_+^2}\bra{\Psi_+^2}_{B_2 C}$. Finally, Bob applies a Bell state measurement $y_2$ on his subsystems, which, for every outcome, produces a maximally entangled state between Alice and Charlie. Thus, we can apply the previously stated observation to conclude that the state $\rho_{AB_1}^{eras}\otimes \rho_{B_2 C}^{eras}$ is nonlocal: in other words, the erased state \eqref{erased} is a nonlocal resource.

It is important to stress that this result is valid for any $k$. This means that there exist states which provide only local correlations in measurement scenarios involving an arbitrary (finite) number of measurements in one of the parties and an infinite number of measurements in the other party, but two copies provide nonlocality in a very simple two-measurement scenario.

\textit{Random two-qubit states.--} We would like to start exploring to which extent \textit{general two-qubit states} are nonlocal resources. Those that violate a Bell inequality certainly are such, so we should concentrate on those that don't. However, at the moment of writing, the existence of a local model for some regime of parameters is known only for very few families of states \cite{werneretal}. Here, we rather adopt a less ambitious scope, and try to see which two-qubit states that \textit{do not violate the CHSH inequality} give rise to nonlocal correlations in the tripartite scenario. A similar restriction was adopted in Refs.~\cite{sen,polish}. In Ref. \cite{miguel tamas} the authors focus on the bipartite scenario and exhibit states $\rho_{1}$ and $\rho_{2}$, such that neither $\rho_{1}^{\otimes N}$ nor $\rho_{2}^{\otimes N}$ violate the CHSH inequality for any $N$, but the state $\rho_{1} \otimes \rho_{2}$ does violate that inequality.

The methods applied are partially based on the fact that every one-way entanglement distillable state is a nonlocal resource: many copies of them violate a Bell inequality in the scenario of Fig.~\ref{Fig1BC}b \cite{us}. A sufficient criterion for a state to be one-way distillable is that its local entropy is greater than its global entropy, \ie,
\ba\label{hashing}
 \textrm{max}\{S\left( \rho_{A} \right),S\left( \rho_{B} \right)\} > S\left( \rho_{AB} \right),
\ea
where $S\left( \rho \right) = - \textrm{Tr}\left( \rho \textrm{log} (\rho) \right)$ is the von Neumann entropy of $\rho$ \cite{hashing}. Note however that, while the previous examples used only two copies of the state, in the present case Alice, Bob, and Charlie must share many copies of $\rho_{AB}$ to obtain nonlocality.  

The algorithm we use in our study works as follows. First, a random two-qubit density matrix is drawn according to the Hilbert-Schmidt measure \cite{Zyc}. Then, we check if the given state violates the CHSH inequality by means of the necessary and sufficient criterion for violation proposed in Ref.~\cite{Horodecki}. If the state does not violate the CHSH inequality, the sufficient criterion \eqref{hashing} is checked: if it is satisfied, the state is indeed a nonlocal resource, even though it does not violate the CHSH inequality. 

We picked $10^{6}$ random states, of which $~99.1\%$ happened not to violate the CHSH inequality. Among these states, we find that $~0.08\%$ are one-way entanglement distillable, and, thus, nonlocal resources. 

Also, we applied the same methods to the states given in \cite{miguel tamas}. Remarkably, we found that both $\rho_{1}$ and $\rho_{2}$ are nonlocal resources according to our criteria, even though neither $\rho_{1}^{\otimes N}$ nor $\rho_{2}^{\otimes N}$ is able to violate the CHSH inequality. %The states we considered can be found at \cite{site}.
%\textbf{VS to RR: had you not found, moreover, that $\rho_1$ and $\rho_2$ of \cite{miguel tamas} are NLR??}

\textit{Two-qubit states under local decoherence.--} In the previous section, we considered randomly picked two-qubit states. However, in practice, one usually deals with specific types of noise such as depolarization, dephasing or amplitude damping \cite{nielsen}. It is thus worthy to study the nonlocality properties of quantum states subjected to these noisy processes. Here we consider the same problem as above described, namely, to test if states that do not violate the CHSH inequality are nonlocal resources, applied to two-qubit states when the mentioned decoherence processes act upon the systems. 

We draw $10^{5}$ two-qubit pure states according to the Fubini-Study measure \cite{Zyc}; then, each of the qubits is evolved \textit{locally} according to the three processes mentioned before. A parameter $t \in [0,1]$ parametrizes the strength of the process, which can also be thought as the time during which the system is under decoherence. For instance, $t = 0$ means that no decoherence has yet acted upon the system, while $t = 1$ means that the system has fully decohered under the process. For definiteness, we assume that both qubits undergo the same type of decoherence and for the same time $t$. For each initial pure state we considered $10^{3}$ equally spaced time-steps in the interval $[0,1]$. Then, in each evolution step, we test for the sufficient criterion for activation of nonlocality described in the previous section. We then compute the number of initial states that present activation of nonlocality at some stage of the evolution, and also the mean number of steps in which activation occured, for each channel. The results are summarized in table \ref{tab}.

\begin{table}
\begin{tabular}{|c|c|c|}
\hline
Process & NLR states (\%) & NLR interval \\
% & states & interval \\
\hline
AD & $92.6$ & $0.078 \pm 0.078$ \\
PD & $68.5$ & $0.023 \pm 0.020$ \\
D & $56.4$ & $0.005 \pm 0.002$ \\
\hline
\end{tabular}
\caption{For each decoherence process (amplitude damping (AD), phase damping (PD), and depolarization (D)), it is presented the percentage of initial states that, on some stage of the evolution, do not violate the CHSH inequality and are nonlocal resources (NLR) according to the criteria presented, and the mean width (over all the initial states) of the evolution interval for which this is observed, with corresponding standard deviations, in units of $t$.}\label{tab}
\end{table}

\begin{figure}
\centering
\includegraphics[width = 0.4\textwidth]{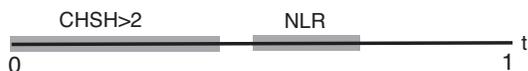}
	\caption{The figure pictorially shows the interval in which the a given evolved state is nonlocal ($CHSH>2$) and do not violate the CHSH inequality but is a nonlocal resource (NLR), for a particular decoherence process, in terms of the parameter $t$.}
\label{Fig2}
\end{figure}

\textit{Discussions.--} We have shown that nonlocality can be extracted from bipartite systems much more easily if they are shared in a tripartite network. For instance Ref. \cite{miguel tamas} showed that there are bipartite states that do not violate the CHSH inequality, but two copies of them do. Here we show that by simply considering three parties it is possible to find examples of states from which an arbitrary finite number of measurements would provide only local correlations, but two copies of the same state shared between three parties provide nonlocality with only two measurements per party. 

The tripartite scenario also provides an interesting link between the nonlocality properties of a given state and its capability of distributing nonlocality when used as a channel in quantum networks. This fact gives additional motivation to seek for robust states, which, in turn, is related to the search for unbounded violations of Bell inequalities. We thus expect that this work will motivate further works on these topics. 

We thank F. Brand\~ao, I. Villanueva, W. D\"ur, and N. Brunner for discussions. This work is supported by the National Research Foundation and the Ministry of Education of Singapore. DC acknowledges the PVE-CAPES program (Brazil).

\textbf{Appendix - Decoherence processes.}
The action of each decoherence process can be described by the map
$\rho^{\prime} = \sum_{i} E_{i} \rho E_{i}^{\dagger}$, where $\rho$ is the initial state of the system \cite{nielsen}.
The strength of the process 
is given by a parameter $t \in [0,1]$, that can be viewed as the probability of full action of the process on the system. Assuming the quantum system is a single qubit, 
the depolarization (D) process are described by
\ba
E^{D}_{0} = \sqrt{1 - \frac{3t}{4}}\openone, \quad E^{D}_{i} = \sqrt{\frac{t}{4}\sigma_{i}},\nonumber
\ea
where $\sigma_{i}$, for $i = 1,2,3$, are the Pauli matrices. For the phase damping (PD) process\ba
E^{PD}_{0} = \sqrt{t}\openone, \quad E^{PD}_{1} = \sqrt{1-{t}}\sigma_3;
\nonumber
\ea
and, for amplitude damping (AD),
\ba
E^{AD}_{0} & = & \ket{0}\bra{0} + \sqrt{1-t}\ket{1}\bra{1}, \quad
E^{AD}_{1}  =  \sqrt{t}\ket{0}\bra{1}.
\nonumber
\ea
\end{document}